\documentclass[twocolumn,superscriptaddress,showpacs,aps,prb]{revtex4}
\usepackage{makeidx}
\usepackage{bm}
\usepackage{graphics}
\usepackage[dvips]{epsfig}
\newcounter{fig}

\begin{document}
\title{Unusual field and temperature dependence of Hall effect in graphene}
\author{L.A. Falkovsky}
\affiliation{L.D. Landau Institute for Theoretical Physics, Moscow
117334, Russia} \affiliation{Institute of the High Pressure
Physics, Troitsk 142190, Russia} 
\begin{abstract}
We calculate   the classic Hall conductivity and mobility
of the undoped and doped (or in the gate voltage)
graphene as a function of temperature, magnetic field, and carrier concentration.
Carrier collisions with defects and acoustic phonons are taken into account.
The Hall resistivity varies almost linearly with temperature.  The
 magnetic field dependence of  resistivity and mobility is anomalous
  in weak magnetic fields.
There is the square root contribution from the field in the
resistivity. The Hall mobility diverges logarithmically with the
field for  low doping.
\end{abstract}
\maketitle
\vskip 5mm

 The widespread attention is attracted to  the recent
investigations \cite{Novo,ZSA} of a mono-atomical  layer  of
graphite with a honeycomb lattice (graphene).  Among the various
reasons for this interest, there are the following. The single
layer of graphite can be wrapped into 0d fullerenes, rolled into
1d nanotubes, and stacked into 3d graphite \cite{SDD}. Hence one
has a possibility to study the dimensionality effects for the
unique substance \cite{Fe}. Graphene has a very simple band
structure in the range of several eV around  the Fermi level. As
has long been shown in study of graphite \cite{W,SW}, the energy
bands of  its single layer are cones
$\varepsilon_{1,2}(\mathbf{p})=\pm vp$ at the corners $ K$ in the
2d Brillouin zone with $v=10^8$ cm/s and the Fermi energy
$\varepsilon_F=0$. So the Fermi surface shrink into points. Such a
degeneration is conditioned by  symmetry because the small group
$C_{3v}$  of the $K$ points has two-dimensional representation.

According to the symmetry consideration, this spectrum (of the
Dirac type but massless and two-dimensional) turns out to be
stable with respect to the Coulomb interaction, what was shown in
Ref. \cite{AB70} for the case of the 3d dimensionality.  Authors
of the recent works \cite{Kh1,SHW} argue, that graphene alters its
behavior under doping from a marginal Fermi liquid to the ordinary
2d Fermi liquid. Graphene exhibits Shubnikov--de Haas oscillations
with the temperature dependence explained in terms of the standard
Fermi-liquid theory. The weak-localization corrections to the
conductivity can have the different sign \cite{SA} depending on
the interaction range of
 impurity potentials or be strongly
suppressed \cite{Kh} due to the Dirac singularity of the spectrum.

Much theoretical efforts \cite{Li,LFS,PGC,TTT,Zk,NMD,Kat} have
been devoted to  evaluate in different approaches the
 minimal conductivity discovered in Refs. \cite{Novo,ZSA}. The finite values of
 conductivity at low temperature means that 2d graphene turns out
 to be a metal (or a
 semimetal) in  contradiction with the recent theoretical analysis
 \cite{AE}.

 The challenging task of the carrier interaction
 with defects in graphene and in the underlying substrate was studied
 in Refs. \cite{PGC,ZA,CF,OGM}. In order to simplify the transport
 problem we
 have considered \cite {FV} the quantum conductivity of graphene in the
 collision-less   limit, when the electric field frequency or
 the spatial dispersion are much important in comparison with the
 collision rate, $(\omega, kv)\gg \tau^{-1}$. We found that the
 conductivity consists of two terms --- one of the Drude type and
 another describing the interband carrier transitions. Since there
 is no a
 gap between the conduction and valence bands, these two terms
 can compete  and the interband contribution becomes  larger at
 high frequencies $\omega>T$. In the opposite case, the intraband contribution plays the
 leading role. Then the quantum considerations give the same
 result as the semiclassical Boltzmann equation.

The Hall effect was studied in Refs.  \cite{ZA,SMJ,HK} in both
quantum and classic regimes for zero temperature,  including the
gap due to interaction-induced phenomena \cite{GSC} or
interactions with chemical adsorbents \cite{HAS}.

In present Letter, we consider the classic Hall effect in both
undoped and doped (or in the gate voltage)  graphene for  finite
temperatures taking into account scattering processes by defects
as well as by acoustic phonons. \vskip 5mm

 Calculations of graphene transport properties meet a fundamental
 difficulty because the modern methods (for instance, the diagrammatic
 approach) are restricted by the  requirement that the mean free path
  $\ell =v\tau $ of
carriers must be much larger than the electron wavelength $\lambda
=h/p_{F}$, i.e., $\ell p_{F}\gg 1$. This condition cannot
evidently be satisfied in the case of undoped graphene where
$p_F=0$. One can avoid this difficulty addressing the problem to
doped samples or to finite temperatures when $T\gg\tau^{-1}$. In
the last case, the temperature appears instead of the Fermi energy
and electrons obey the Boltzmann statistics.

 At low temperatures,  collisions with defect are important.
   Using the  Fermi golden rule, one can find the
 collision rate for scattering by defects,
 \begin{equation}\label{gr}
\tau^{-1}_{imp}(\varepsilon)=\frac {n_{imp}}{2\pi} \int d^2\mathbf
{p'}|u(\mathbf {p-p'})|^2(1-\cos{\theta})
\delta[\varepsilon-\varepsilon(\bf{p'})],
\end{equation}
  where $n_{imp}$ is the defect concentration per the  unit
  surface
and $u(\bf{p-p'})$ is the Fourie component of the defect
potential. As  noticed previously, we are interested in the
carrier energy of  the order of  temperature $\varepsilon\simeq
T$, whereas  the defect potential varies on interatomic distances.
Therefore,  the potential in  integrand  (\ref{gr}) for the
intravalley scattering     can be considered as a constant $u({\bf
p-p'})=u_0\simeq \varepsilon_0a^2$ ($u_0$ and $a$ are the energy
and the distance of the atomic scale), and  $\cos{\theta}$
disappears in integrating over the angle. But for  the intervalley
processes, carriers view  the neighboring valley in  small angles
and that gives a small factor $Ta/v$. Then the intervalley
scattering can be ignored. Integration (\ref{gr}) gives
\[
\tau^{-1}_{imp}(\varepsilon)= n_{d}|\varepsilon|,\]
 where
$n_{d}=n_{imp}(u_0/v)^2$ is the defect concentration per lattice
unit. It is important that the collision rate becomes proportional
to the energy. Since the electron density of states is also
proportional to the energy,  one obtains the residual resistance
independent of temperature.

The temperature dependence arises due to scattering by phonons.
  At low temperatures, the electron collisions with acoustic phonons
are essential. Extending  the result of Ref. \cite{FF} to the 2d
electron system, we obtain
\begin{equation}
\tau _{el-ph}^{-1}=\alpha |\varepsilon|T/T_{D},\label{pc}
\end{equation}
 where $\alpha$ is a constant of the order of the unity and
$T_D$ is the Dedye temperature (around   2000 K for graphene). For
the important values of energy $\varepsilon\simeq T$,  Eq.
(\ref{pc}) gives $\tau^{-1}\simeq T^2/T_D$ in comparison with
$\tau^{-1}\simeq T^3/T_D^2$ for  3d systems. Notice, that all
scattering angles are essential now, since the Fermi surface (or
the chemical potential) is assumed to be small.

As the processes of electron-defect and electron-phonon scattering
are independent,  the total scattering rate can be written in the
form
\[
\tau^{-1}=\tau^{-1}_{imp}(\varepsilon)+\tau^{-1}_{el-ph}(\varepsilon)=
|\varepsilon|n_d^*, \]
where the notation
 \[n_d^*=n_d+\alpha T/T_D\]
is introduced.
\vskip 5mm

The solution to the Boltzmann equation  in  the $\tau
$-approximation results in the electrical conductivity
\[\sigma_{\alpha \beta}=-\frac{e^2v^2}{\pi}\sum_{bands} \int
d\varepsilon
\frac{df_0}{d\varepsilon}\frac{m\tau}{1+(\Omega\tau)^2}\left(
\begin{array}{cc}
1 & \Omega\tau \\
-\Omega\tau & 1
\end{array}
\right) ,
\]
where the factor 4 is acquired  due to summation over spin and two
$K$ points per the Brillouin zone. The integration is performed
over $\varepsilon$
 from 0 to $\infty$ in the conduction band and  from $-\infty$ to 0
 in the valency band,
the cyclotron mass  $m=\varepsilon/v^2$, the cyclotron frequency
$\Omega=eH/mc$, the magnetic field $H$ is assumed to be normal to
the graphen layer, and $f_0(\varepsilon-\mu)$ is the Fermi
function.

 As the result, the Hall conductivity tensor  takes the form
\begin{equation}
\sigma_{xx}=\frac{e^2}{4\pi\hbar n_d^*}\int_{-\infty}^{\infty}
\frac{\varepsilon^4 d\varepsilon}{\varepsilon^4+\eta^2}
\text{sech}^{2}{\frac{\varepsilon-\varphi}{2}}, \label{dia}
\end{equation}
\begin{equation}
\sigma_{xy}=\frac{e^2\eta}{4\pi\hbar n_d^*}\int_{-\infty}^{\infty}
\frac{\varepsilon^2\text {sign}(\varepsilon)
d\varepsilon}{\varepsilon^4+\eta^2}
\text{sech}^{2}{\frac{\varepsilon-\varphi}{2}}, \label{ndia}
\end{equation}
where we restore the Planck constant  and introduce the
dimensionless $\eta$ magnetic field and the dimensionless chemical
potential $\varphi$:
\[\eta=|e|\hbar Hv^2/cn_d^*k_B^2T^2, \quad \varphi=\mu/k_BT.
\]

From measurements, longitudinal resistivity $\rho$ and  Hall angle
$\theta_H$ (or  Hall mobility $\mu_H$) can be obtained:
\[\rho=\frac{\sigma_{xx}}{\sigma^2_{xx}+\sigma^2_{xy}},\,
\tan\theta_H=\frac{\sigma_{xy}}{\sigma_{xx}},\,
\mu_H=\frac{\sigma_{xy}}{H\sigma_{xx}}.
\]
 \begin{figure}[h]
\resizebox{.5\textwidth}{!}{\includegraphics{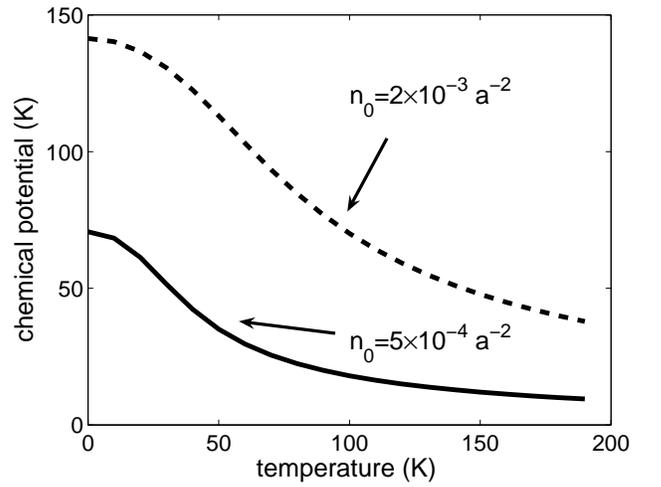}}
\caption{Chemical potential versus temperature for two doping
values  $n_0$ per the lattice unit. } \label{resdep1}
\end{figure}
 The chemical potential $\mu=0$ for undoped graphene and in the absence
 of the gate voltage. In the field effect experiment, i.e.  under
 the gate voltage (or for the doped material), $\mu$ is determined
 by the fixed carrier concentration $n_0$ :
\[n_0=\frac{2}{\pi v^2}\int_0^{\infty}\varepsilon
[f_0(\varepsilon-\mu)-f_0(\varepsilon+\mu)]d\varepsilon.
\]
As  seen from Fig. \ref{resdep1}, while
  the temperature grows, the chemical potential
  tends to its value $\mu=0$ in the undoped graphene.
\vskip 5mm

One can see from Eqs. (\ref{dia}) -- (\ref{ndia}), that as it must
the diagonal component of conductivity is an even function of the
chemical potential and the off-diagonal one is an odd function.

\begin{figure}[h]
\resizebox{.5\textwidth}{!}{\includegraphics{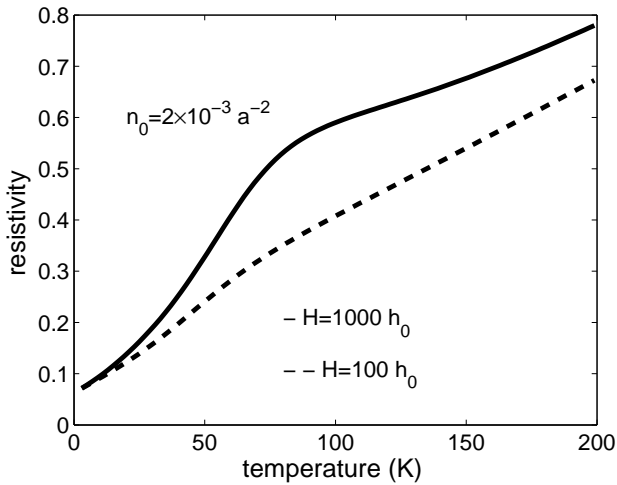}}
\caption{Resistivity (in units $\hbar/e^2$) versus temperature for
two values of the magnetic field H (in units $h_0=ck_B^2K^2/e\hbar
v^2=0.11 $ G) and the doping level $n_0$ indicated in Fig. The
electron collisions with uncharged defects (of concentration 0.02
per the lattice unit here and everywhere in Figs.) and  acoustic
phonons are taken into account. } \label{resdep2}
\end{figure}

 The expansion   of conductivity tensor components
  in terms of magnetic field ($\eta\ll 1$) for a case
of nondegenerate carrier statistic ($\varphi\ll 1$) takes a form
\begin{equation}
\begin{array}{c}
\sigma_{xx}=\sigma_0[1-(\pi/4\sqrt{2})\sqrt{\eta}]\, ,\\
\sigma_{xy}=0.25\sigma_0\eta\varphi\ln{(1/\eta)}\, ,
\end{array}
\label{wf}
\end{equation}
where $\sigma_0$ is the conductivity  in zero  field,
\[\sigma_0=\frac{e^2}{\pi \hbar n_d^*}\, .
\]
 The resistivity in the absence of field grows linearly with temperature
 (see Fig. \ref{resdep2}) and  is independent of the chemical potential
 $\mu $, i. e., of  doping.

\begin{figure}[h]
\resizebox{.5\textwidth}{!}{\includegraphics{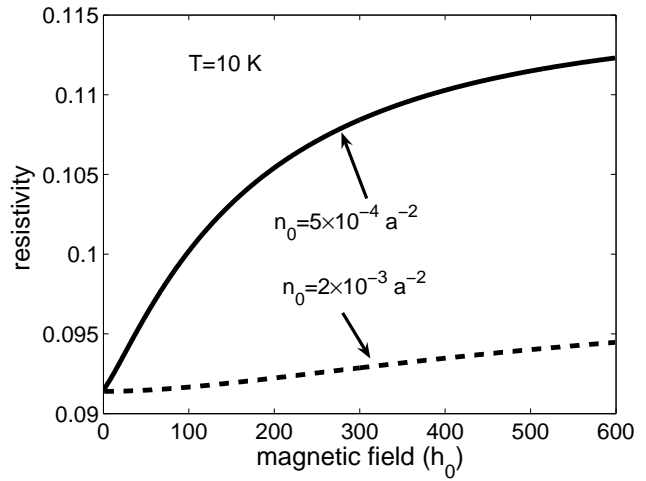}} \caption{
Resistivity (in units $\hbar/e^2$) as a function of the magnetic
field (in units $h_0=0.11 $ G) at given temperature and doping
$n_0$ labeled at the curves. } \label{resdep3}
\end{figure}

Let us underline that the resistivity corresponding to
conductivity (\ref{wf}) grows as the square root of the magnetic
field  (also see Fig. \ref{resdep3}, the upper curve) in contrast
to the ordinary squared field dependence.

\begin{figure}[h]
\resizebox{.5\textwidth}{!}{\includegraphics{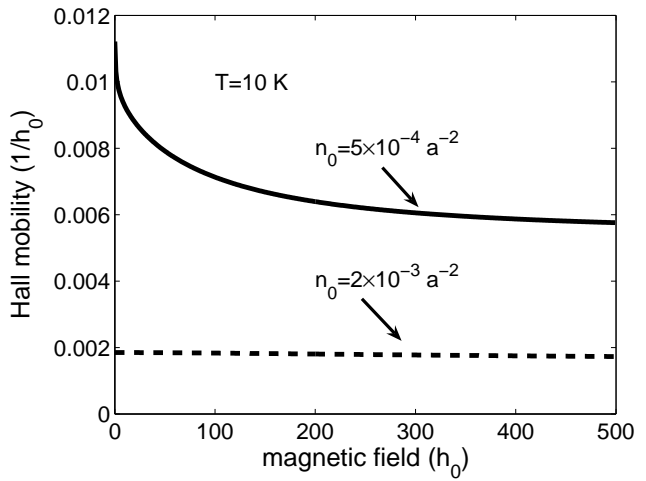}} \caption{
 Hall mobility (in units $1/h_0$) as a function of magnetic field
(in units $h_0=0.11 $ G) for two doping $n_0$ labeled at the
curves. } \label{resdep4}
\end{figure}

As  mentioned above, the Hall field $\sigma_{xy}$ does not vanish
only in the doped material and  alters the sign if the sign of the
chemical potential $\mu=\varphi T$ becomes changed, while
undergoing from holes to electrons. In weak fields, the Hall field
as well as longitudinal conductivity
  reveals the anomalous field dependence --
   the logarithmic divergence  of  the Hall mobility
 (see Fig. \ref{resdep4}, the upper curve).
\begin{figure}[h]
\resizebox{.5\textwidth}{!}{\includegraphics{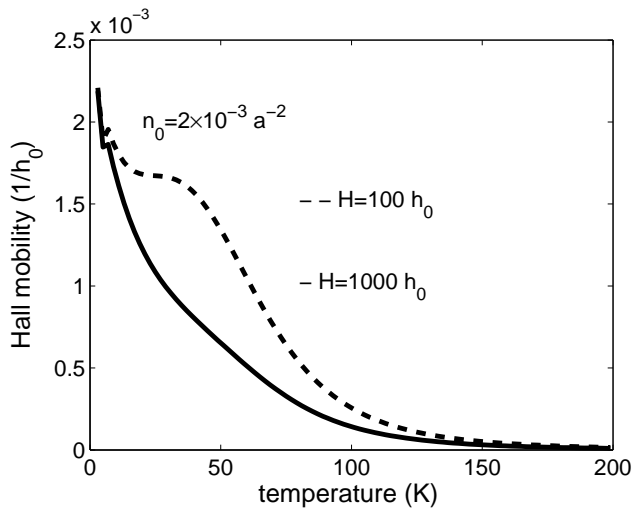}}
\caption{Hall mobility (in units $1/h_0$) versus temperature for
two values of magnetic field $H$ (in units $h_0=0.11 $ G). }
\label{resdep5}
\end{figure}
The rise of Hall mobility is cut in weak fields at a size of the
sample  $r$, which must be larger than the Larmor radius
$r_c=v/\Omega$. This condition restricts the field, $\eta>\hbar
v/rk_BTn_d^*$.

For the  nondegenerate carrier statistic ($\varphi\ll 1$) but in
the opposite limit of  high fields ($\eta\gg 1$), the conductivity
tensor has the form:
\begin{equation}
\begin{array}{c}
\sigma_{xx}=\sigma_0[(7/15)\pi^4+1.23\times 2^4\varphi^2]/\eta^2\, ,\\
\sigma_{xy}=2.77\sigma_0\varphi/\eta\, .
\end{array}
\label{dia1}
\end{equation}
  In high
fields, $\sigma_{xy}$ is inversely proportional to the field (see
the second line in Eq. (\ref{dia1})), i.e. the Hall coefficient in
high fields is indeed independent of the field. As in  high fields
$\sigma_{xx}\propto H^{-2}$,
 the longitudinal resistivity  tends to "saturate"
 (see Fig. \ref{resdep3}).

For the degenerate statistic ($\varphi\gg 1$), we obtain from Eqs.
 (\ref{dia}) -- (\ref{ndia})
\begin{equation}
\begin{array}{c}
\sigma_{xx}=\sigma_0\varphi^4/(\eta^2+\varphi^4)\, ,\\
\sigma_{xy}=\sigma_0\eta\varphi^2/(\eta^2+\varphi^4)\, .
\end{array}
\label{dst}
\end{equation}
As expected, these equations result, first, in the resistivity
$\rho=1/\sigma_0$ independent of the field (see the lower curve in
Fig. \ref{resdep3}) and, second, in the Hall coefficient
$R=1/ecn_0$ which gives the carrier concentration $n_0$.

The Hall mobility as a function of temperature is shown in Fig.
\ref{resdep5}. In  relatively high magnetic fields, the mobility
 is governed by resistivity, but  it has more complicated behavior
 in weak fields.
\vskip 6mm

{\it In conclusions}, considering electrons and holes in graphene
as ordinary Fermi liquids, we found that the longitudinal
resistivity varies mainly linearly with temperature.  Hall
resistivity and mobility vary anomalously in weak magnetic fields
for a case of low carrier concentrations while the degenerate
statistic holds. There is the square root contribution from the
field in the resistivity. The Hall mobility diverges
logarithmically with the field for the low doping. This anomalous
behavior results from the linear dependence of both
 electron density of state and collision rate for the Dirac fermions in
graphite.

 This work was supported by the Russian
Foundation for Basic
Research.  

\end{document}